\documentclass[english]{article}
\usepackage[T1]{fontenc}
\usepackage[latin9]{inputenc}
\usepackage{geometry}
\usepackage{setspace}
\usepackage{babel}
\begin{document}

\title{On the Security of Two Blind Quantum Computations}

\author{Shih-Min Hung and Tzonelih Hwang%
\thanks{Corresponding Author, email: hwangtl@ismail.csie.ncku.edu.tw%
}}
\maketitle
\begin{abstract}
Blind quantum computation (BQC) protocol allows a client having partial
quantum ability to delegate his quantum computation to a remote quantum
server without leaking any information about the input, the output
and the intended computation to the server. Several BQC protocols
have been proposed, e.g., Li et al. in {[}1{]} proposed a triple-server
BQC protocol and Xu et al. in {[}2{]} proposed a single-server BQC
protocol. Though both papers claimed that their protocols can satisfy
the requirement of privacy, this paper points out a security loophole
in their protocols. With that the server can reveal the private information
of the client.\end{abstract}
\begin{description}
\item [{Keywords:}] Blind Quantum Computation; Quantum Cryptography.
\end{description}

\section{Introduction}

Blind quantum computation (BQC) is one of the most important research
topics in quantum cryptography that enables a client to delegate a
quantum computation to a quantum server without revealing any information
about the input, the output and the intended computation to the server.
Since Childs {[}3{]} proposed the first BQC protocol in 2005, a variety
of BQC protocols have been proposed {[}4-8{]}.

In 2009, Broadbent et al. {[}5{]} presented the first universal single-server
BQC protocol, and also proposed a double-server BQC protocol modified
from the single-server BQC protocol. In the double-server BQC protocol,
the client can be completely classical if both servers (Bob1 and Bob2)
pre-share entangled states and do not communicate with each other.
Recently, Li et al. {[}1{]} claimed that the restrictions of non-communicating
servers and pre-sharing of entangled states between both servers can
be removed if one more server is introduced. Hence, they proposed
a triple-server BQC protocol based on the technique of entanglement
swapping. In their protocol, three servers can communicate with one
another and the client can be almost classical, i.e., only with the
capability of receiving and sending qubits. Nevertheless, Xu et al.
{[}2{]} later indicated that it is unnecessary for the client to communicate
with three servers, and hence proposed a single-server BQC protocol. 

However, this paper will point out a loophole in both Li et al. and
Xu et al.\textquoteright{}s protocols. With this loophole, the server
is able to reveal the private information, such are the input, the
output and the intended computation, which the client does not want
the server to know. In this paper, we will use Xu et al.\textquoteright{}s
single-server BQC protocol as an example to describe the loophole
and the attack. 

The rest of this article is organized as follows. Section 2 reviews
Xu et al.\textquoteright{}s single-server BQC protocol. Section 3
describes the attack on Xu et al.\textquoteright{}s protocols. Finally,
a concluding remark is given in Section 4.

\section{Review Xu et al.\textquoteright{}s single-server BQC protocol}

Suppose that a client Alice with limited quantum capability wants
to delegate a quantum problem to a quantum server Bob without revealing
any information about the input, the output and the intended computation
with the help of a trusted center, Charlie, who helps them to generate
the $m$-qubit graph states. Xu et al.\textquoteright{}s single-server
protocol proceeds as follows.
\begin{description}
\item [{Step$\,$1.}] Charlie generates $4m$ Bell pairs $\left|\psi_{0,0}\left(B_{k},A_{k}\right)\right\rangle =\frac{1}{\sqrt{2}}\left(\left|00\right\rangle +\left|11\right\rangle \right)\left(k=1,2,...,4m\right)$
and distributes the particles $B_{k}$ of all Bell pairs to Bob, the
other particles $A_{k}$ to Alice.
\item [{Step$\,$2.}] Alice randomly selects $2m$ particles $A_{s_{1}},A_{s_{2}},...A_{s_{m}}$
and $A_{t_{1}},A_{t_{2}},...A_{t_{m}}$, where $1\leq s_{i}\leq2m<t_{i}\leq4m$,
$i\in\left\{ 1,2,...,m\right\} $ from $A_{k}$ and sends them to
Bob. She discards the others.
\item [{Step$\,$3.}] After Bob receives these $2m$ particles form Alice,
he implements Bell measurement on $\left(A_{s_{i}},A_{t_{i}}\right)$
and transmits the measurement outcome of $\left|\psi_{z_{i}',x_{i}'}\left(A_{s_{i}},A_{t_{i}}\right)\right\rangle $
to Alice, where $\left(z_{i}',x_{i}'\right)\in\left\{ 0,1\right\} ^{2}$
and $i\in\left\{ 1,2,...,m\right\} $.
\item [{Step$\,$4.}] By the measurement outcome of $\left|\psi_{z_{i}',x_{i}'}\left(A_{s_{i}},A_{t_{i}}\right)\right\rangle $
and the entanglement swapping, Alice gets the combined state of the
corresponding particles $B_{s_{i}}$ and $B_{t_{i}}$ , $\left|\psi_{z_{i}',x_{i}'}\left(B_{s_{i}},B_{t_{i}}\right)\right\rangle \left(i=1,2,...,m\right)$,
which are the particles that Bob gets from Charlie.
\item [{Step$\,$5.}] Alice sends $2m$ classical message $\left\{ \tilde{\theta}_{k}=\left(-1\right)^{x_{k}}\theta_{k}+z_{k}\pi\right\} _{k=1}^{2m}$
to Bob, where $\left\{ \theta_{s_{i}}\right\} _{i=1}^{m}$ are randomly
selected from $S=\left\{ k\pi/4|k=0,1,...,7\right\} $ and $\left\{ \left(z_{s_{i}},x_{s_{i}}\right)\right\} _{i=1}^{m}$
depend on $\left\{ \left(z_{s_{i}}',x_{s_{i}}'\right)\right\} _{i=1}^{m}$
in Step 4. The other $\theta_{k}$ and $\left(z_{k},x_{k}\right)$
are selected randomly.
\item [{Step$\,$6.}] Bob measures his first $2m$ particles in the basis
$\left\{ \pm\tilde{\theta}_{k}\right\} _{k=1}^{2m}$ and sends the
measurement results $\left\{ b_{k}\right\} _{k=1}^{2m}$ to Alice.
\item [{Step$\,$7.}] Upon receiving $\left\{ b_{k}\right\} _{k=1}^{2m}$
form Bob, Alice keeps $\left\{ b_{s_{i}}\right\} _{i=1}^{m}$ and
discards the others and subsequently sends the classical information
$\left\{ t_{i}\right\} _{i=1}^{m}$ to Bob. Bob keeps the particles
$\left\{ B_{t_{i}}\right\} _{i=1}^{m}$ and labels them as $\left\{ B_{i}\right\} _{i=1}^{m}$
in order. Hence, Bob has the $m$ qubit graph state $\otimes_{i=1}^{m}\left|\theta_{s_{i}}+b_{s_{i}}\pi\right\rangle $.
\item [{Step$\,$8.}] Since Bob has the $m$ qubit graph state $\otimes_{i=1}^{m}\left|\theta_{s_{i}}+b_{s_{i}}\pi\right\rangle $
and only Alice knows the values of $\theta_{s_{i}}$ and $b_{s_{i}}$,
Alice can run Broadbent et al.\textquoteright{}s single-server BQC
protocol to delegate the quantum problem to Bob.
\end{description}
In Xu et al.\textquoteright{}s protocol, Alice only has to receive
the photons and resend it to Bob, whereas in the original Broadbent
et al.\textquoteright{}s protocol, Alice has to generate single photon
and rotate it. It seems that the client in Xu et al.\textquoteright{}s
protocol requires less quantum ability than in Broadbent et al.\textquoteright{}s
protocol. However, in the following, we shall point out that Xu et
al.\textquoteright{}s protocol is not as secure as it claimed to be.

\section{Server\textquoteright{}s attack on Xu et al.\textquoteright{}s BQC
protocol}

In this section, we show that the server (Bob) can obtain the secret
information of the client (Alice) without being detected in Xu et
al.\textquoteright{}s BQC protocol. 

As we know, in Broadbent et al.\textquoteright{}s single-server BQC
protocol, the security of the private information is based on $\left|\theta_{i}\right\rangle =\left|0\right\rangle +e^{i\theta_{i}}\left|1\right\rangle \left(i=1,2,...,m\right)$,
where only Alice knows the $\theta$ of each particle. Hence, Bob
cannot disclose the input, the output and the intended computation
without knowing $\theta$. Similarly, the security of the private
information in Xu et al.\textquoteright{}s BQC protocol is also based
on $\left|\theta_{i}\right\rangle =\left|0\right\rangle +e^{i\theta_{i}}\left|1\right\rangle \left(i=1,2,...,m\right)$.
In their protocol, Xu et al. claimed that no one except Alice knows
the positions ( $\left\{ s_{i}\right\} _{i=1}^{m}$ and $\left\{ t_{i}\right\} _{i=1}^{m}$
). And Bob cannot calculate without having and . It also means that
the input, the output and the intended computation will not be revealed. 

However, in this article, we will point out two ways for Bob to obtain
$\left\{ s_{i}\right\} _{i=1}^{m}$ and $\left\{ t_{i}\right\} _{i=1}^{m}$
in Xu et al.\textquoteright{}s BQC protocol. The first attack is an
eavesdropping attack. Inside the protocol, because Alice does not
have quantum memory, Charlie has to send the particles to Alice one
by one. When Alice receives a particle, she has to decide to resend
it or discard it at that time. If Bob eavesdrops the quantum channel
between Alice and Charlie, then Bob will know the positions ( $\left\{ s_{i}\right\} _{i=1}^{m}$
and $\left\{ t_{i}\right\} _{i=1}^{m}$ ) Alice selected to resend.

For example, when Charlie sends the first particle to Alice, if Alice
resends it to Bob, Bob will get that particle at that time and he
will know that Alice resends it; otherwise, if Alice discards it,
Bob will not get any particle at that time and he will know that Alice
discards it. 

The second attack is a Trojan horse attack {[}9-11{]}. Since the quantum
bits are transmitted twice in these protocols, Bob can insert invisible
photons or delay photons to the photons sent from Charlie to Alice.
Then, Bob can measure the invisible photons or delay photons sent
from Alice to Bob and obtain the positions ( $\left\{ s_{i}\right\} _{i=1}^{m}$
and $\left\{ t_{i}\right\} _{i=1}^{m}$ ). In these two attacks, Bob
can get the secret $\left\{ s_{i}\right\} _{i=1}^{m}$ and $\left\{ t_{i}\right\} _{i=1}^{m}$
easily and then calculate the input, the output and the intended computation.
These problems also can be found in Li et al.\textquoteright{}s BQC
protocol.

\section{Conclusions}

We have shown that Xu et al.\textquoteright{}s BQC protocol scheme
is not secure against server\textquoteright{}s attack. A server can
obtain client\textquoteright{}s information without being detected.
The same attack can also be successful in Li et al.\textquoteright{}s
BQC. The Trojan horse attack can be easily prevented by device. However,
the eavesdropping attack be an interesting future research.

\section*{Acknowledgment }

We would like to thank the Ministry of Science and Technology of Republic
of China for financial support of this research under Contract No.
MOST 104-2221-E-006-102 -.

\end{document}